\documentclass[11pt]{article}
\usepackage[normalem]{ulem}
\usepackage{graphicx,amssymb,amsmath,amsfonts,empheq,latexsym,braket,bm,bbm,dsfont,upgreek}
\usepackage{macro}

\title{
Note on explicit construction of conformal generators on the fuzzy sphere
}

\author[1]{Ruihua Fan}

\affil[1]{\normalsize\it Department of Physics, University of California, Berkeley, CA 94720, USA}

\begin{document}

\maketitle	

\begin{abstract}
The lowest Landau level on the sphere was recently proposed as a continuum regularization of the three-dimensional conformal field theories, the so-called fuzzy sphere regularization.
In this note, we propose an explicit construction of the conformal generators on the fuzzy sphere in terms of the microscopic Hamiltonian.
Specifically, we construct the generators for the translation and special conformal transformation, which are used in defining the conformal primary states and thus are of special interest.
We apply our method to a concrete example, the fuzzy sphere regularized three-dimensional Ising conformal field theory. We show that it can help capture all primaries with spin $\ell < 4$ and scaling dimension $\Delta < 7$.
In particular, our method can clearly separate the primary from other states that differ in scaling dimension by $1\%$, making it hard otherwise based solely on using the conformal tower associated with the primaries.

\end{abstract}

\tableofcontents

\newpage

\section{Introduction}

Understanding (2+1)-dimensional conformal field theory (CFT) is one of the central challenges in theoretical physics~\cite{DiFrancesco:1997nk,Rychkov:2016iqz}.
Its strongly correlated nature necessitates using non-perturbative approaches to extract information reliably, such as numerical simulation of microscopic models and the conformal bootstrap~\cite{Simmons-Duffin:2016gjk,Poland:2018epd}.
Recent developments offer a new numerical method, the so-called fuzzy sphere regularization~\cite{Madore:1991bw,Hasebe:2010vp,Zhu:2022gjc,Zhou:2023qfi,O(3)FuzzySphere:2023,Hofmann:2023llr}.
This method models CFTs by confining multiple flavors of electrons to the lowest Landau level (LLL) and producing critical fluctuations in charge-neutral flavor degrees of freedom. Thus, it enjoys having a finite-dimensional Hilbert space and continuous spatial symmetries simultaneously~\cite{Sondhi1993,Zaletel:2018HalfFilledLL}.
By placing the system on a two-dimensional sphere~\cite{Haldane:1983xm,Greiter2011}, the state-operator correspondence ensures a one-to-one mapping between the local operators in the target CFT and energy eigenstates at the critical point.
Therefore straightforward diagonalization of the microscopic Hamiltonian is able to reveal various properties of the CFT~\cite{Hu:2023xak, Han:2023yyb, Hu:2024pen, Zhou:2023fqu, Hu:2023ghk,Cuomo:2024psk,Zhou:2024dbt,Dedushenko:2024nwi}.

In applying this method broadly, two challenges arise.
The first is locating a critical point that minimizes finite-size effects~\cite{Lao:2023zis}. 
This work is to address the second problem, distinguishing primary states in the spectrum.
To explain it, recall that the defining feature of a CFT is having the conformal algebra, which includes the translations $P_\mu$, (spacetime) rotations $J_{\mu\nu}$, dilation $D$ and the special conformal transformations (SCT) $K_\mu$. 
The Hilbert space on the two-dimensional sphere is organized into a direct sum of infinitely many irreducible representations of the conformal algebra, also called the conformal families.
Each family is labeled by a unique member, the primary state, that is annihilated by the SCT generator $K_\mu$. 
We can obtain the rest of the family, the descendant states, by applying translation operators $P_\mu$ repeatedly to the primary.
Therefore, these primary states encode all the dynamical information of the CFT and are of special interest.

Previous numerical studies of fuzzy sphere regularization identify the primary states by comparing the spectrum from exact diagonalization with results from the conformal bootstrap.
However, finite-size corrections might lead to incorrect identification especially when eigenenergies are closely spaced.
On the other hand, having access to the SCT generators $K_\mu$ allows us to identify the primary states faithfully, providing us with a more systematic approach.
In this work, we propose an explicit construction of the translation and SCT generators on the fuzzy sphere in terms of the microscopic degrees of freedom. 
Our construction is not meant to be exact at all energy scales but to approximate the true conformal generators only in the low-energy and small-spin regimes.
As a concrete example, we apply our general formula to the fuzzy sphere regularized Ising CFT and identify all low-energy primary states with spin $\ell < 4$ and scaling dimensions $\Delta < 7$.
The obtained result here is consistent with that in the previous study~\cite{Zhu:2022gjc}.

\section{Review of the fuzzy sphere regularization}
\label{sec:review}

Consider the Landau level problem on a 2D sphere of radius $R$ in the unit $|e| = \hbar = c = 1$~\cite{Haldane:1983xm,Greiter2011}.
Placing a magnetic monopole of charge $q>0$ at the center of the sphere, the number of fluxes is $N_\phi = 2q \in \bbZ$.
The LLL contains $N_\phi + 1$ orbitals and the wavefunctions are given by the monopole harmonics $\psi_{q,m}(\Omega)$, $m=-N_\phi/2,\ldots,N_\phi/2$ (see \appref{app:review of LL problem} for a review).
We consider $N\geq 2$ flavors of electrons and label them by $a = 1,\ldots,N$. In general, we will fix the system at an integer or rational filling where the charge-neutral flavor degrees of freedom can undergo a quantum phase transition while the charged fluctuations remain gapped.

Let $c_{a}(\Omega)$ denote the LLL-projected fermion annihilation operator in the real space and $c_{a,m}$ in the orbital space, which are related by 
\begin{equation}
    c_a(\Omega) = \sum_{m = -q}^q \psi_{q,m}(\Omega) c_{a,m}\,.
\end{equation}
The Hamiltonian generally contains two parts $H = H_1 + H_2$, the one-body term 
\begin{equation}
    H_1 = M_{ab} \int c_{a}^\dag(\Omega) c_{b}(\Omega) d\Omega
\end{equation}
and the two-body interaction
\begin{equation}
	H_2 = \int V_{abcd}(\Omega_{12}) c_{a}^\dag(\Omega_1) c_b^\dag(\Omega_2) c_c^\dag(\Omega_2) c_d(\Omega_1) d\Omega_1 d\Omega_2\,.
 \label{eq:twobody}
\end{equation}
where the summation over repeated indices is assumed.
Three-body interactions are necessary for realizing more exotic phase transitions and will be ignored here for simplicitly.
As is discussed in the quantum Hall literature, it is convenient to expand the interaction $V_{abcd}(\Omega_{12})$ in terms of the pseudopotentials~\cite{Haldane:1983xm,KievlsonPseudopotential}.
In practice, keeping the first few terms in the expansion is sufficient to realize a critical point with a small finite-size correction, and we have
\begin{equation}
\label{eq:Vabcd}
	V_{abcd}(\Omega_{12}) = \sum_{k = 0}^{k_{\text{max}}} g_{k,abcd} \nabla_2^{2k} \delta^{(2)}(\Omega_1 - \Omega_2)\,.
\end{equation}
The system has a built-in continuous spatial rotation symmetry. We can construct the rotation generators simply by adding the angular momentum operators for each electron together. 
At the critical point, the system acquires the emergent conformal symmetry and it is in principle possible to obtain other conformal generators as well.
For example, the Hamiltonian itself $H$ is related to the dilation generator.
In the following, we explain how to use conformal algebra to write down microscopic expressions for the translation and SCT generators.

\section{Conformal kinematics}
\label{sec:conformal kinematics}

Given a conformal Killing vector $v_\varepsilon = \varepsilon^\mu(x) \partial_\mu$, the corresponding generator in the quantum theory is a topological surface operator $Q_\varepsilon(\Sigma)$ defined on a co-dimension-1 surface $\Sigma$
\begin{equation}
	\label{eq:conformal charge}
	Q_\varepsilon(\Sigma) = \int_\Sigma d\Sigma n^\mu(x) \varepsilon^{\nu}(x) T_{\mu\nu}(x)\,,
\end{equation}
where $T_{\mu\nu}(x)$ is the stress-energy tensor.
In many-body systems defined by the microscopic Hamiltonians, only the temporal component $T_{00}$, related to the Hamiltonian density, is naturally accessible. 
Although we can formally define other components as well, the numerical implementation will be difficult in practice.  
Therefore, our construction involves only $T_{00}$.

In the $d$-dimensional Euclidean spacetime $\bbR^d$, the conformal Killing vectors for the dilation, translation and SCT are 
\begin{equation}
    v_D = x^\mu \partial_\mu\,,\quad 
    v_{P_\mu} = \partial_\mu\,,\quad
    v_{K_\mu} = - x^2 \partial_\mu + 2 x_\mu x^\nu \partial_\nu\,.
\end{equation}
Let $r$ denote the radial coordinate and $\Omega$ the angular part. We perform a coordinate transformation $x_\mu \mapsto (\tau,\Omega)$, $r = e^{\tau}$ and move to the cylindrical geometry $\bbR\times S^{d-1}$, with $\bbR$ being the imaginary time direction. 
In the new coordinate, the conformal killing vectors on the unit sphere $x^2 = 1$ read
\begin{equation}
	v_D = \partial_\tau\,,\quad v_{P_\mu} + v_{K_\mu} = 2 x^\mu \partial_\tau\,,\quad x^2 = 1\,.
\end{equation}
where $x_{\mu=1,\ldots,d}$ is the $\mu$-th component of the unit vector in $\bbR^d$.
It follows from \eqref{eq:conformal charge} that the corresponding quantum operator is
\begin{equation}
\label{eq:D X formula}
	D = \int_{S^{d-1}} T_{00}(\Omega) d\Omega\,,\quad P_\mu + K_\mu = 2\int_{S^{d-1}} x_\mu T_{00}(\Omega) d\Omega
\end{equation}
where $T_{00}(\Omega)$ is the stress-energy tensor on the unit sphere.
Recalling the conformal algebra $[D,P_\mu] = P_\mu$, $[D,K_\mu] = -K_\mu$, we have
\begin{equation}
	P_\mu - K_\mu = [D, P_\mu + K_\mu]\,.
\end{equation}
Introduce the notation $X_\mu = (P_\mu + K_\mu)/2 = \int_{S^{d-1}} x_\mu T_{00} d\Omega$. We can write the translation and SCT generators as
\begin{equation}
\label{eq:P K formula}
	P_\mu = X_\mu + [D, X_\mu]\,,\quad 
	K_\mu = X_\mu - [D, X_\mu]\,.
\end{equation}
The right-hand side of \eqnref{eq:P K formula} only contains $T_{00}(\Omega)$ as we have shown in \eqnref{eq:D X formula}.

\section{Implementation on the fuzzy sphere}
\label{sec:implementation}

\subsection{General formulation}

Suppose that the system is at the conformally invariant critical point. 
In the thermodynamic limit, the Hamiltonian $H$ should be proportional to the dilation generator $D$ up to an additive constant, i.e. $D = \alpha H$. In a finite-size system, irrelevant operators can also occur on the left-hand side and will be ignored in our analysis. 
Here $\alpha$ is a non-universal number and can be determined by normalizing the spectrum. 
Specifically, we normalize it so that the ground state has zero energy and the spin-2 state, which corresponds to the stress-energy tensor, has an energy of 3.
It follows that the Hamiltonian density $h(\Omega)$ is proportional to the stress-energy tensor $T_{00}(\Omega)$ up to a total derivative
\begin{equation}
    T_{00}(\Omega) = \alpha h(\Omega) + \partial O(\Omega)\,.
\end{equation}
The total derivative term $\partial O(\Omega)$, also in the form of an irrelevant local operator, reflects the ambiguity in the Hamiltonian density for generic interactions and cannot be determined a priori~\cite{Kitaev:2005hzj, Kapustin:thermalhall}.\footnote{We would like to thank Yin-Chen He for pointing this out for us.} 
When there are only finite terms in the pseudopotential expansion \eqnref{eq:Vabcd}, we can adopt the following canonical form for the Hamiltonian density
\begin{equation}
	h(\Omega_1) = M_{ab} c_{a}^\dag(\Omega_1) c_{b}(\Omega_1) + \int V_{abcd}(\Omega_{12}) c_{a}^\dag(\Omega_1) c_b^\dag(\Omega_2) c_c^\dag(\Omega_2) c_d(\Omega_1) d\Omega_2\,.
\end{equation}
Since the total derivative $\partial O$ is irrelevant, it should be less important when the system size is sufficiently large. 
However, it may have a non-negligible effect for intermediate system sizes.
For simplicity, we adopt the above definition for $h(\Omega)$ and assume $\partial O(\Omega) = 0$ to derive explicit expressions for the conformal generators $P_\mu$ and $K_\mu$ in terms of the microscopic parameters. 
We will discuss this assumption further later.

The essential part is to calculate $X_\mu = (P_\mu + K_\mu)/2$. 
Given the fact that the numerical implementation of the fuzzy sphere usually conserves angular momentum $\ell_z$, it is convenient to focus on $X_z$ which preserves the rotation symmetry around the $z$ axis. The other two generators $X_{x}$ and $X_y$ can be similarly constructed through the formalism below. The rest of this section sketches the main idea of the calculation. Readers primarily interested in the concrete example may proceed directly to the next section.

Specifically, we can also separate $X_z = X_{z,1} + X_{z,2}$ into a one-body and two-body term
\begin{equation}
\begin{aligned}
	X_{z,1} =& \alpha M_{ab} \int \cos\theta c_{a}^\dag(\Omega) c_{b}(\Omega) d\Omega\,, \\
	X_{z,2} =& \alpha \int \cos\theta_1 V_{abcd}(\Omega_{12}) c_{a}^\dag(\Omega_1) c_b^\dag(\Omega_2) c_c^\dag(\Omega_2) c_d(\Omega_1) d\Omega_1 d\Omega_2\,.
\end{aligned}
\end{equation}
While it is possible to directly substitute the lowest Landau level (LLL) wave functions and perform a brute-force calculation, it becomes much more difficult for the two-body term.
Here we choose a more systematic approach using algebraic properties of the monopole harmonics. 
For the one-body term $X_{z,1}$, we expand the electron operators in the orbital space, yielding
\begin{equation}
    X_{z,1} = \alpha M_{ab} \sum_{m=-q}^q \int \cos\theta \, \psi_{q,m}^*(\Omega) \psi_{q,m}(\Omega) d\Omega\, c_{a,m}^\dag c_{b,m} \,.
\end{equation}
We recognize $\cos\theta$ as the monopole harmonics $\sqrt{4\pi/3} Y_{0,1,0}(\Omega)$ and can directly use the property of the monopole harmonics (see \eqnref{eq: thm 2} of \appref{app:review of LL problem}).
We have
\begin{equation}
    X_{z,1} = \alpha M_{ab} \sum_{m=-q}^q \frac{m}{q+1} c_{a,m}^\dag c_{b,m} \,.
\end{equation}
For the two-body term, we first plug in  \eqnref{eq:Vabcd} to expand the interaction potential $V_{abcd}(\Omega_{12})$ with respect to the real-space pseudopotentials and have
\begin{equation}
\begin{gathered}
    X_{z,2} = \alpha \sum_{k=0}^{k_{\text{max}}} g_{k,abcd} I^{(k)}_{m_1,m_2;m_3,m_4} c_{a,m_1}^\dag c_{b,m_2}^\dag c_{c,m_3} c_{d,m_4} \\
    I^{(k)}_{m_1,m_2;m_3,m_4} = \int \cos\theta_1 \Big(\nabla_2^{2k} \delta(\Omega_{12}) \Big) \psi_{q,m_1}^*(\Omega_1) \psi_{q,m_2}^*(\Omega_2) \psi_{q,m_3}(\Omega_2) \psi_{q,m_4}(\Omega_1) d\Omega_1 d\Omega_2
\end{gathered}
\end{equation}
where the angular momentum conservation $m_1+m_2 = m_3+m_4$ is assumed implicitly. 
We can further expand the contact interaction in terms of the spherical harmonics as follows
\begin{equation}
	\nabla^{2k} \delta(\Omega_{12}) = \sum_{\ell=0}^{+\infty} \sum_{m = -\ell}^{\ell} (-\ell(\ell+1))^k Y_{0,\ell,m}^*(\Omega_1) Y_{0,\ell,m}(\Omega_2)
\end{equation}
which allows us to decompose the double integral $I^{(k)}_{m_1,m_2;m_3,m_4}$ into two independent single integrals over $\Omega_1$ and $\Omega_2$.
Each of these single integrals can then be computed analytically in the same way as the one-body term.
We refer the reader to \appref{app:details} for further details.

\subsection{Example: Ising CFT}

Let us apply our general formula to a concrete example, the fuzzy-sphere regularized Ising CFT~\cite{Zhu:2022gjc}.
Specifically, consider the bilayer quantum Hall problem of spinless electrons at the unit total filling. We label the two layers (the flavors of the electrons) by $a=\, \uparrow,\downarrow$. 
The Hamiltonian takes a similar form as the transverse-field Ising model
\begin{equation}
\label{eq:Ising fuzzy}
\begin{aligned}
	H = - h\int  n_x(\Omega) d \Omega + \int V(\Omega_{12}) n_\uparrow (\Omega_1) n_\downarrow(\Omega_2) d \Omega_1 d \Omega_2  
\end{aligned}
\end{equation}
where $n_x = c_\uparrow^\dag c_\downarrow + c_\downarrow^\dag c_\uparrow$ favors the interlayer coherence and $V(\Omega_{12})$ is a repulsive interaction that favors a layer-ferromagnetic order.
Equivalent to \eqnref{eq:Vabcd}, we use the Haldane pseudopotentials $V_{J}$'s to parameterize the interaction potential $V(\Omega_{12})$ (see \appref{app:converting}). 
The system exhibits the Ising criticality at $h/V_1 = 3.16$, $V_0/V_1 = 4.75$, $V_{J>1} = 0$ with an extremely small finite-size effect~\cite{Zhu:2022gjc}.

\begin{figure}
\centering
\includegraphics[width=6.3cm]{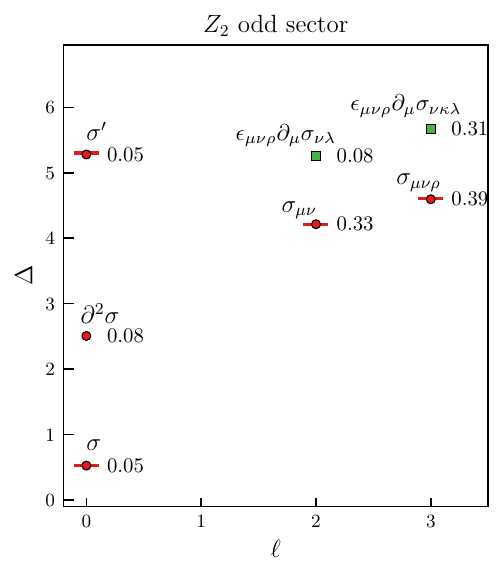}
\hspace{10pt}
\includegraphics[width=6.3cm]{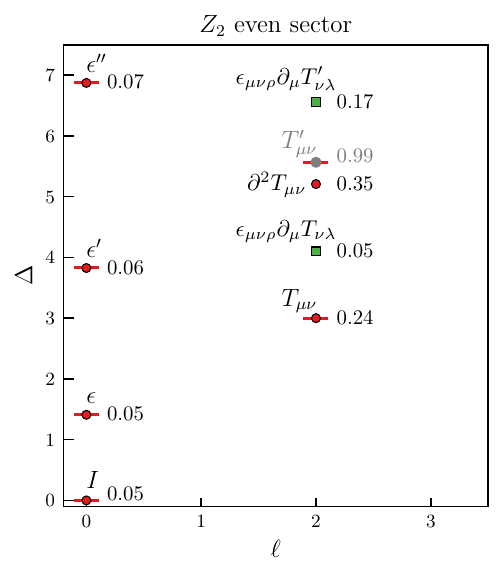}
\caption{Norm square of $\tilde{K}_z \ket{\psi}$ for the energy eigenstates $\ket{\psi}$. Numbers are the numerical values. Red dots are states with the even parity and green squares are ones with the odd parity. Red lines are what we expect from the CFT. Some descendant states turn out to yield small $\calN_{K\psi}$ due to its scaling dimension being special ($\partial^2\sigma$), conformal algebra ($\epsilon_{\mu\nu\rho}\partial_\mu O_{\nu\ldots}$ for a spinful primary $O_{\nu\ldots}$) or conservation law ($\partial^2 T_{\mu\nu}$). The spin-2 primary $T_{\mu\nu}'$, denoted by the gray dot and text, shows a relatively large $\calN_{K\psi} \approx 0.99$ that exceeds the cutoff.}
\label{fig:Kpsi norm}
\end{figure}

We then follow the recipe explained in the previous sections to construct the putative SCT generator $\tilde{K}_z$. Here we add a tilde to emphasize that $\tilde{K}_z$ is microscopically defined and should be distinguished from $K_z$ in the CFT. Then, we can determine whether an energy eigenstate $\ket{\psi}$ is a primary state by computing
\begin{equation}
    \calN_{K\psi} \equiv \Vert \tilde{K}_z \ket{\psi} \Vert^2
\end{equation}
Namely, $\ket{\psi}$ being a primary state implies that $\calN_{K\psi}$ vanishes. In the numerical calculation, we expect to see a sufficiently small number due to the finite-size correction.

We choose a system of $N=12$ electrons and compute $\calN_{K\psi}$ for states $\ket{\psi}$ with spin $\ell < 4$ and rescaled energy $\Delta < 7$ in the $\ell_z = 0$ sector.
\figref{fig:Kpsi norm} shows all states for which $\calN_{K\psi} < 0.4$. 
This cutoff is chosen based on a gap in the norms, as the next smallest value is $\calN_{K\psi} \approx 0.62$.
This criterion correctly selects all primary states in the spin-$\bbZ_2$ odd sector, as well as all scalar primaries and the stress-energy tensor in the $\bbZ_2$ even sector.
The result provides a useful guidance of identifying the primaries in the spectrum.
For example, there are two $\bbZ_2$-odd spin-2 states with close energies 4.21 and 4.41, which should correspond to a primary $\sigma_{\mu\nu}$ and a descendant $\partial_\mu\partial_\nu \partial^2 \sigma$.
The state at 4.21 has a significantly smaller $\calN_{K\psi} \approx 0.33$ compared to the other $\calN_{K\psi} \approx 5.7$, which allows us to assign $\sigma_{\mu\nu}$ to the state at 4.21 confidently.
Another example involves the three $\bbZ_2$-even spin-2 states at energies $5.509$, $5.566$ and $5.815$, which are candidates of one primary $T_{\mu\nu}'$ and two descendants $\partial_\mu \partial_\nu \partial^2 \varepsilon$, $\partial_\mu \partial_\nu \varepsilon'$.
We find $\calN_{K\psi} \approx 7.32, 0.99$ and $9.630$, respectively, indicating that the one at energy $5.566$ should correspond to the primary $T_{\mu\nu}'$ even though its $\calN_{K\psi} \approx 0.99$ exceeds the cutoff.

Additionally, \figref{fig:Kpsi norm} contains a few descendant states showing small values of $\calN_\psi$. We can understand their presence from exact CFT calculations.
As for $\partial^2 \sigma$, this is because $\Delta_{\sigma} \approx 0.518$ giving
\begin{equation}
    \frac{\braket{\partial^2 \sigma|K_z^\dag K_z |\partial^2 \sigma}}{\braket{\partial^2 \sigma|\partial^2 \sigma}} = 2(2\Delta - 1) / 3 \approx 0.024
\end{equation}
which is small even in the CFT.
The state $K_z \ket{\partial^2 T_{\mu\nu}}$ also has a parametrically small norm in the $\ell_z = 0$ sector, as $T_{\mu\nu}$ is a conserved current.
As for the four descendants $\epsilon_{\mu\nu\rho}\partial_\mu\sigma_{\nu\lambda}$, $\epsilon_{\mu\nu\rho}\partial_\mu\sigma_{\nu\kappa\lambda}$, $\epsilon_{\mu\nu\rho}\partial_\mu T_{\nu\lambda}$ and $\epsilon_{\mu\nu\rho}\partial_\mu T_{\nu\lambda}'$ (green squares in the plot), applying $K_z$ to their $\ell_z = 0$ components yields identically zero results.

\begin{figure}
\centering
\includegraphics[width=5.6cm]{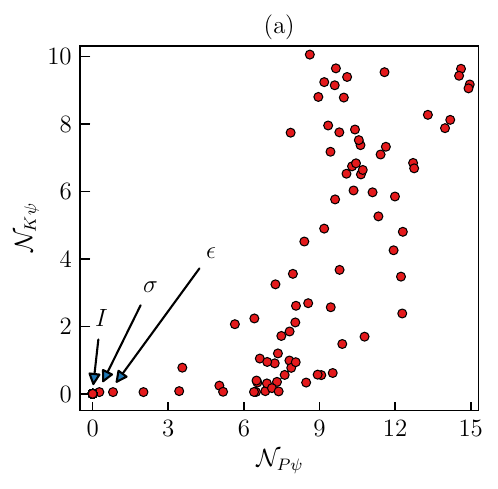}
\hspace{10pt}
\includegraphics[width=6cm]{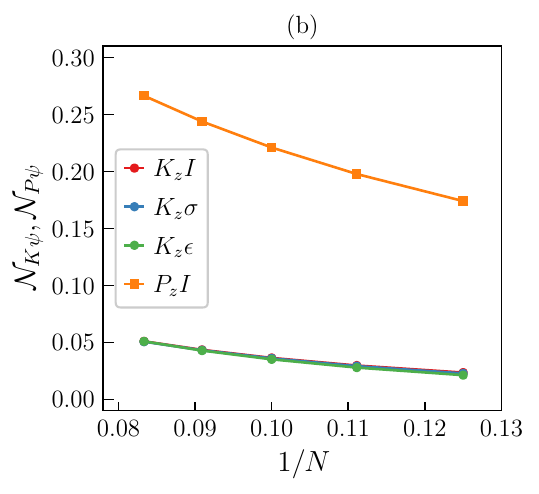}
\caption{(a) Scatter plot of $\calN_{P\psi}$ and $\calN_{K\psi}$ for the low-energy states. The computed $\calN_{P\psi}$ for the first three primary states $I$, $\sigma$, $\epsilon$ are 0.266, 0.808, 2.017, respectively. The result for $\calN_{P\psi}$ shows a relatively large deviation from the theoretical expectation compared with that for $\calN_{K\psi}$. (b) Finite size scaling of the computed norm $\calN_{P\psi}$ and $\calN_{K\psi}$. We fix $h,V_0,V_1$ and change the system size from $N=8$ to $N=12$. The red, blue, and green lines overlap strongly.}
\label{fig:Kpsi Ppsi FSS}
\end{figure}

Our recipe produces the translation generator $\tilde{P}_z$ at the same time and we can similarly compute the norm
\begin{equation}
    \calN_{P\psi} \equiv \Vert \tilde{P}_z \ket{\psi} \Vert^2
\end{equation}
which should vanish only for the conformally invariant ground state.
\figref{fig:Kpsi Ppsi FSS}~(a) shows a scattered plot of the computed norms $(\calN_{P\psi}, \calN_{K\psi})$ for the first $\lesssim 150$ low-energy states. 
The ground state has the smallest $\calN_{P\psi} \approx 0.27$ and stays closest to the origin.
The first excited state $\sigma$ has $\calN_{P\psi} \approx 0.81$, comparable to the theoretical expectation $2\Delta_\sigma \approx 1.036$, and thus is next closest to the origin.
Interestingly, we find that the constructed SCT generator $\tilde{K}_z$ performs better than the translation generator $\tilde{P}_z$ for unknown reasons. 
For example, the red and orange curves in \figref{fig:Kpsi Ppsi FSS}~(b) show $\calN_{K\psi}$ and $\calN_{P\psi}$ for the ground state.
We can see that $\calN_{K\psi}$ is much smaller whereas both quantities are expected to be zero in the CFT.

\section{Discussion}

In this work, we explicitly construct the conformal generators for the translation and special conformal transformation on the fuzzy sphere.
Practically, having microscopic conformal generators $\tilde{P}_z$ and $\tilde{K}_z$ provides an independent approach to identify a primary state and its descendants. 
Conceptually, it also verifies whether a microscopic Hamiltonian is able to faithfully represents a CFT in the fuzzy sphere regularization.

A crucial assumption in our construction is that the canonically defined microscopic Hamiltonian density $h(\Omega)$ at the critical point is proportional to the stress-energy tensor $T_{00}(\Omega)$ up to some additive constant.
However, it is difficult to know a priori the validity of this assumption.
In the case of fuzzy sphere regularized Ising CFT, we indeed observe discrepancy between our numerical result and the theoretical expectation.
In particular, $\calN_{K\psi}$ for the descendant states and almost all results for $\calN_{P\psi}$ have a large deviation from the theoretical values that can be computed via conformal algebra.
\figref{fig:Kpsi Ppsi FSS}~(b) shows the finite-size scaling of the computed $\calN_{K\psi}$ for the first three primary states and $\calN_{P\psi}$ for the ground states. We can see that all curves monotonically increase instead of decrease. 
Moreover, the proposed method here fails to probe any spin-4 primary. 
One possibility is that the parameter in the Hamiltonian has a small deviation from the actual critical point, resulting in a relevant perturbation that enters the constructed generators.
It is also possible that the actual stress-energy tensor $T_{00}(\Omega)$ differs from the Hamiltonian density $h(\Omega)$ by some total derivative terms that become important in intermediately larger system sizes that are not close to the thermodynamic limit yet~\cite{Fardelli:2024qla}.
Understanding this deviation more systematically can help improve our method.
It would also be valuable to explore alternative methods for constructing the conformal generators, such as those based on entanglement~\cite{Lin:2023pvl}, to avoid the ambiguities inherent in the current method.

\emph{Note added:} When completing our manuscript, we notice Ref.~\cite{Fardelli:2024qla} which has overlaps with the current manuscript.

\section*{Acknowledgment}

We are grateful to Yin-Chen He for helpful discussions that inspire the current work.
We thank Junkai Dong and Ashvin Vishwanath for collaborations on related projects, and providing valuable feedback to the current manuscript. 
R.F. is supported by the Gordon and Betty Moore Foundation.

\appendix

\section{Review of the Landau level problem on the sphere}
\label{app:review of LL problem}

The Landau level problem on the sphere can be solved analytically and the solution of the single-particle wavefunctions are known as the Wu-Yang monopole harmonics~\cite{Wu:1976ge,Wu:1977qk}.
In this section, we review some basic facts of this problem and collect useful properties of the monopole harmonics~\cite{Greiter2011}.

Consider an electron of charge $e = -|e| = -1$ and mass $M$ moving on the sphere of radius $R$. A magnetic monopole of charge $q$ sits at the center of the sphere and generates a radial magnetic field
\begin{equation}
    \bm{B} = \frac{q}{r^2} \bm{e}_r\,.
\end{equation}
The total magnetic flux through the sphere is $\Phi_{tot} = 2|q| \phi_0$, with $\phi_0 = 2\pi \hbar c/|e| = 2\pi$ the flux quantum. We require $2q \in \bbZ$ to have a well-defined quantum mechanical problem~\cite{dirac1931quantised}.
The Hamiltonian only has the kinetic term and reads 
\begin{equation}
    H = \frac{1}{2MR^2} \Lambda^2 = \frac{\omega_c}{2q} \Lambda^2\,,\quad
    \bm{\Lambda} = \bm{r} \times (-i \bm{\nabla} + \bm{A})
\end{equation}
where $\omega_c = B/M$ is the cyclotron frequency and $\bm{\Lambda}$ is the dynamical angular momentum operator of the \emph{electron}. Without loss of generality, we choose a gauge $\bm{e}_r \cdot \bm{A} = 0$ so that the electron angular momentum has no radial component, $\bm{e}_r \cdot \bm{\Lambda} = \bm{\Lambda} \cdot \bm{e}_r = 0$.
However, $\bm{\Lambda}$ does not satisfy the $\su(2)$ Lie algebra but the following one
\begin{equation}
    [\Lambda_i, \Lambda_j] = i \varepsilon_{ijk} (\Lambda_k - q e_{r,k})\,,\quad i=x,y,z
\end{equation}
Therefore, it cannot be identified as the generator of rotation.
Instead, we need to include contributions from the electromagnetic field and consider the \emph{total} angular momentum
\begin{equation}
    \bm{L} = \bm{\Lambda} + q \bm{e}_r\,,
\end{equation}
where $q \bm{e}_r$ is the angular momentum of the electromagnetic field created by the electron and the magnetic monopole.
Note that our definition of $q$ differs from that in Ref.~\cite{Wu:1976ge,Wu:1977qk} by a minus sign because of setting $e = -|e| = -1$.
One can show that
\begin{equation}
    [L_i, X_j] = i \varepsilon_{ijk} X_k\,,\quad
    \bm{X} = \bm{e}_r,\bm{\Lambda},\bm{L}
\end{equation}
In particular, we see that $\bm{L}$ satisfies the correct $\su(2)$ Lie algebra and the Hilbert space should form the corresponding unitary representation.
One can get intuition by focusing on the south (or north) pole and taking the limit $R\rightarrow \infty$, where the sphere becomes an infinite plane. In this limit, $L_{\theta}$ and $L_\phi$ are reduced to the guiding center momenta. 

The $\su(2)$ algebra completely determines the Landau level structure.
Noticing that $\bm{L} \cdot \bm{e}_r = \bm{e}_r \cdot \bm{L} = q$, we have
\begin{equation}
    \Lambda^2 = L^2 - q^2 \geq 0\,.
\end{equation}
Combining $L^2 = \ell(\ell+1)$, $2\ell=0,1,2,\ldots$ in an $\su(2)$ irreducible representation and $2q \in \bbZ$, one can deduce
\begin{equation}
    \ell = |q| + n\,,\quad n = 0,1,2,\ldots
\end{equation}
Thus, different irreducible representations of the $\su(2)$ correspond to different Landau levels. 
The eigen energy of the $n$-th Landau level is
\begin{equation}
    E_n = \omega_c \left[ \big( n + \frac{1}{2} \big) + \frac{n(n+1)}{2q} \right]\,,\quad n = 0,1,2,\ldots
\end{equation}
The degeneracy is determined by the dimension of the corresponding irreducible representation
\begin{equation}
    N_{\text{$n$-th LL}} = 2|q| + 2n + 1\,.
\end{equation}
The extra $2n + 1$ term is related to the ``shift" of the quantum Hall problem~\cite{Wen:1992ej}.

The eigenstates, known as the monopole harmonics $Y_{q,\ell,m}$, is a simultaneous eigenstate of the total angular momentum $L^2$ and $L_z$ 
\begin{equation}
    L^2 Y_{q,\ell,m} = \ell (\ell + 1) Y_{q,\ell,m} \,,\quad
    L_z Y_{q,\ell,m} = m Y_{q,\ell,m}\,.
\end{equation}
The expression of $L^2$ and $L_z$ depends on the gauge choice, and so does $Y_{q,\ell,m}$.
One common choice is the latitudinal gauge $\bm{A} = -(q/R)\cot \theta \bm{e}_\varphi$. In this case, the monopole harmonics for the lowest Landau level, denoted by $\psi_{q,m}$, take a simple form. For $q>0$, we have
\begin{equation}
    \psi_{q,m}(\theta,\phi) = \sqrt{\frac{(2q+1)!}{4\pi (q+m)!(q-m)!}} \cos^{q+m}\frac{\theta}{2} \sin^{q-m}\frac{\theta}{2} e^{im\phi}\,,\quad m = -q,\ldots,|q|\,.
\end{equation}
Comparing our notation with that in Ref.~\cite{Wu:1976ge,Wu:1977qk}, we have $\psi_{q,m} = Y_{-q,q,m}$ up to a gauge transformation.
The monopole harmonics is reduced to the ordinary spherical harmonics at $q=0$.
Many properties of the spherical harmonics can be generalized to the monopole harmonics.
We collect some useful gauge-independent properties below
\begin{itemize}
\item Orthonormal relation
\begin{equation}
\label{eq:orthonormal relation}
    \int d\Omega Y_{q,\ell,m} Y_{q,\ell',m'}^* = \int_0^{2\pi} d\phi \int_0^\pi d\theta \sin\theta  Y_{q,\ell,m} Y_{q,\ell',m'}^* = \delta_{\ell,\ell'} \delta_{m,m'}
\end{equation}
\item Complex conjugation
\begin{equation}
\label{eq:complex conjugation}
    Y_{q,\ell,m}^* = (-1)^{q+m} Y_{-q,\ell,-m}
\end{equation}
\item If $q+q'+q'' = 0$ and $m+m'+m'' = 0$, we have
\begin{equation}
\begin{aligned}
	Y_{q,\ell,m} & Y_{q',\ell',m'} = \\
	\sum_{\ell''} & (-1)^{\ell+\ell'+\ell''} \sqrt{\frac{(2\ell+1)(2\ell'+1)(2\ell''+1)}{4\pi}} \begin{pmatrix} \ell & \ell' & \ell'' \\ m & m' & m'' \end{pmatrix} \begin{pmatrix} \ell & \ell' & \ell'' \\ q & q' & q'' \end{pmatrix} Y^*_{q'', \ell'', m''}
\end{aligned}
\label{eq: thm 1}
\end{equation}
where we have used the Wigner $3j$ symbol.
\item It follows from \eqnref{eq:orthonormal relation} and \eqref{eq: thm 1} that
\begin{equation}
\begin{aligned}
    \int Y_{q,\ell,m} & Y_{q',\ell',m'} Y_{q'',\ell'',m''} d\Omega \\
    & = (-1)^{\ell+\ell'+\ell''} \sqrt{\frac{(2\ell+1)(2\ell'+1)(2\ell''+1)}{4\pi}} \begin{pmatrix} \ell & \ell' & \ell'' \\ m & m' & m'' \end{pmatrix} \begin{pmatrix} \ell & \ell' & \ell'' \\ q & q' & q'' \end{pmatrix} 
\end{aligned}
\label{eq: thm 2}
\end{equation}
\end{itemize}

\section{Details of the integral}
\label{app:details}

In \secref{sec:implementation}, we have shown that the coefficient $I^{(k)}_{m_1,m_2;m_3,m_4}$ in the two-body term $X_{z,2}$ can be written in terms of two single integrals. More explicitly, we have
\begin{equation}
\begin{aligned}
    I^{(k)}_{m_1,m_2;m_3,m_4} =&  \sum_{\ell=0}^{+\infty} \sum_{m=-\ell}^\ell (-\ell(\ell+1))^k A^{(k)}_{m_1,m_4} (m,\ell) B^{(k)}_{m_2,m_3} (m,\ell)\,\\
    A^{(k)}_{m_1,m_4} (m,\ell) =& \int \cos\theta Y_{0,\ell,m}^*(\Omega) \psi_{q,m_1}^*(\Omega) \psi_{q,m_4}(\Omega) d\Omega\,, \\
    B^{(k)}_{m_2,m_3} (m,\ell) =& \int Y_{0,\ell,m}(\Omega) \psi_{q,m_2}^*(\Omega) \psi_{q,m_3}(\Omega) d\Omega\,\,.
\end{aligned}
\end{equation}
Here we use the properties of the monopole harmonics above to compute the two integrals $A^{(k)}_{m_1,m_4} (m,\ell)$ and $B^{(k)}_{m_2,m_3} (m,\ell)$ explicitly.

The integral $B^{(k)}_{m_2,m_3}(m,\ell)$ is readily computed via \eqnref{eq: thm 2}, which yields
\begin{equation}
    B^{(k)}_{m_2,m_3} (m,\ell) = (-1)^{q + \ell + m_2} \sqrt{\frac{(2q+1)^2 (2\ell+1)}{4\pi}} \begin{pmatrix} q & q & \ell \\ -m_2 & m_3 & m \end{pmatrix} \begin{pmatrix} q & q & \ell \\ q & -q & 0 \end{pmatrix} 
\end{equation}
Here the Wigner $3j$ symbol is nonzero for $m=m_2 - m_3$, $2q \leq \ell$.
To compute $A^{(k)}_{m_1,m_4} (m,\ell)$, we first apply \eqnref{eq: thm 1} to reduce the number of monopole harmonics 
$$
    \cos\theta Y_{0,\ell,m}^* = \sqrt{\frac{(\ell-m)(\ell+m)}{(2\ell-1)(2\ell+1)}} Y_{0,\ell-1,m}^* + \sqrt{\frac{(\ell+1-m)(\ell+1+m)}{(2\ell+1)(2\ell+3)}} Y_{0,\ell+1,m}^*
$$
Then we can apply \eqnref{eq: thm 2} to each term and have
\begin{equation}
\begin{aligned}
    A^{(k)}_{m_1,m_4} (m,\ell) = & (-1)^{q+\ell+m_4+1} \sqrt{\frac{(2q+1)^2}{4\pi(2\ell+1)}}  \\
    &\quad \times \Big( \sqrt{\ell^2 - m^2} \begin{pmatrix} q & q & \ell-1 \\ -m_1 & m_4 & m \end{pmatrix} \begin{pmatrix} q & q & \ell-1 \\ q & -q & 0 \end{pmatrix} \\
    & \quad \quad \quad  + \sqrt{(\ell+1)^2 - m^2} \begin{pmatrix} q & q & \ell+1 \\ -m_1 & m_4 & m \end{pmatrix} \begin{pmatrix} q & q & \ell+1 \\ q & -q & 0 \end{pmatrix} \Big) 
\end{aligned}
\end{equation}
where $m = m_2 - m_1$.

\section{Converting the Haldane pseudopotentials into real-space pseudopotentials}
\label{app:converting}

There are two complete bases of pseudopotentials that can be used to expand a generic two-body interaction potential $V(\Omega_{12})$.
The first is defined in real space using the delta function and its derivatives 
\begin{equation}
	V(\Omega_{12}) = \sum_{k = 0}^{k_{\text{max}}} g_{k} \nabla_2^{2k} \delta^{(2)}(\Omega_1 - \Omega_2)\,,
\end{equation}
which is the basis we use to construct the translation and SCT generators explicitly. The second basis, known as the Haldane pseudopotential, is defined in angular momentum space:
\begin{equation}
	V(\Omega_{12}) = \sum_{L=0}^{L_{max}} V_{2q-L} P_L(\bm{L}_1 + \bm{L}_2)\,,
\end{equation}
where $P_{L}(\bm{L}_1 + \bm{L}_2)$ is a projector that projects onto the two-particle states with the total angular momentum $L$.
The Haldane pseudopotential is more commonly used in the quantum Hall literature.

To convert between these two expansions, we first express the interaction potential as a series in Legendre polynomials:
\begin{equation}
	V(\Omega_{12}) = \sum_{\ell=0}^{+\infty} U_{\ell} P_\ell(\cos\theta_{12}) \,.
\end{equation}
The relation between the Haldane pseudopotential coefficients $V_{2q-L}$ and $U_\ell$ is given by~\cite{wooten2013haldane}
\begin{equation}
\label{eq:conversion}
	V_{2q-L} = \sum_{\ell=0}^{2q} U_{\ell} (-1)^{2q+L} (2q+1)^2 \begin{Bmatrix} L & q & q \\ \ell & q & q \end{Bmatrix} \begin{pmatrix} q & \ell & q \\ -q & 0 & q \end{pmatrix}^2\,,
\end{equation}
where $\{\cdots\}$ is the Wigner $6j$ symbol.
For the real-space pseudopotential $\nabla^{2k} \delta^{(2)}(\Omega_{12})$, we can straightforwardly compute $4\pi U_\ell = (-\ell(\ell+1))^k (2\ell+1)$, and then apply \eqnref{eq:conversion} to perform the conversion. We will now give the result for the first four pseudopotentials
\begin{itemize}
\item Having only $g_0$
\begin{equation}
	\frac{4\pi V_J}{g_0} = \frac{(2q+1)^2}{4q+1} \delta_{J,0} 
\end{equation}
\item Having only $g_1$ 
\begin{equation}
	\frac{4\pi V_J}{g_1} = -\frac{q(2q+1)^2}{4q+1} \delta_{J,0} + \frac{q(2q+1)^2}{4q-1} \delta_{J,1} 
\end{equation}
\item Having only $g_2$ 
\begin{equation}
	\frac{4\pi V_J}{g_2} = \frac{8 q^3 (2 q+1)^2}{16 q^2-1} \delta_{J,0} -\frac{4 q^2 (2 q+1)^2}{4 q-1} \delta_{J,1} + \frac{4 q^2 (2 q-1) (2 q+1)^2}{16 q^2-16 q+3} \delta_{J,2}
\end{equation}
\item Having only $g_3$ 
\begin{equation}
\begin{aligned}
	\frac{4\pi V_J}{g_3} = & -\frac{4 q^3 (2 q+1)^2 (6 q+1)}{16 q^2-1} \delta_{J,0} + \frac{4 q^3 (2 q+1)^2 (18 q-13)}{16 q^2-16 q+3} \delta_{J,1} \\ & \quad -\frac{4 q^2 (2 q+1)^2 \left(18 q^2-11 q+1\right)}{16 q^2-16 q+3} \delta_{J,2} + \frac{12 (q-1) q^2 (2 q-1) (2 q+1)^2}{(4 q-5) (4 q-3)} \delta_{J,3} 
\end{aligned}
\end{equation}
\end{itemize}
The contact interaction with higher derivatives yields a faster growing $q$ factor when it is converted to the pseudopotential coefficient. Thus, a reasonable thermodynamic limit requires us to consider the following form of the real-space pseudopotential 
\begin{equation}
	\frac{g_0}{q} \delta(\Omega_1 - \Omega_2) + \frac{g_1}{q^2} \nabla^2\delta(\Omega_1 - \Omega_2) + \frac{g_2}{q^3} \nabla^4\delta(\Omega_1 - \Omega_2) + \frac{g_3}{q^4} \nabla^6\delta(\Omega_1 - \Omega_2)
\end{equation}
Then we have
\begin{equation}
	\begin{pmatrix} V_0 \\ V_1 \\ V_2 \\ V_3  \end{pmatrix} = 
	\frac{(2q+1)^2}{4\pi q}
	\begin{pmatrix} 
	\frac{1}{(4q+1)} & -\frac{1}{(4q+1)} & \frac{8q}{16q^2-1} & -\frac{4(6q+1)}{(16q^2-1)} \\
	0 & \frac{1}{(4q-1)} & -\frac{4}{(4q-1)} & \frac{4(18q-13)}{(4q-3)(4q-1)} \\
	& & \frac{4(2q-1)}{(4q-3)(4q-1)} & - \frac{4(2q-1)(9q-1)}{q(4q-3)(4q-1)}\\ 
	& & & \frac{12(q-1)(2q-1)}{q(4q-5)(4q-3)}
	\end{pmatrix}
	\begin{pmatrix} g_0 \\ g_1 \\ g_2 \\ g_3 \end{pmatrix}
\end{equation}

\bibliography{ref.bib}

\end{document}